\documentclass[nofootinbib,aps,superscriptaddress,preprint,floatfix]{revtex4}
\usepackage{epsfig}
\usepackage{setspace}
\usepackage{color}
\usepackage{subfigure}
\usepackage{graphicx}
\usepackage{epstopdf}
\usepackage{axodraw4j}
\usepackage{pstricks}
\usepackage{slashed}

\def\beq{\begin{equation}}
\def\eeq{\end{equation}}
\def\bea{\begin{eqnarray}}
\def\eea{\end{eqnarray}}
\def\beqa{\begin{eqnarray}}
\def\eeqa{\end{eqnarray}}

\begin{document}
\vspace{3.0cm}
\preprint{\vbox {
\hbox{WSU--HEP--1201} 
}}

\vspace*{2cm}

\title{\boldmath Searching for super-WIMPs in leptonic heavy meson decays}

\author{Y. G. Aditya\vspace{5pt}}
\email{ygaditya@wayne.edu}
\affiliation{Department of Physics and Astronomy\\[-6pt]
        Wayne State University, Detroit, MI 48201}

\author{Kristopher J. Healey\vspace{5pt}}
\email{healey@wayne.edu}
\affiliation{Department of Physics and Astronomy\\[-6pt]
        Wayne State University, Detroit, MI 48201}

\author{Alexey A.\ Petrov\vspace{5pt}}
\email{apetrov@wayne.edu}
\affiliation{Department of Physics and Astronomy\\[-6pt]
        Wayne State University, Detroit, MI 48201}
\affiliation{Michigan Center for Theoretical Physics\\[-6pt]
University of Michigan, Ann Arbor, MI 48109\\[-6pt] $\phantom{}$ }

\begin{abstract}
We study constraints on the models of bosonic super-weakly interacting particle (super-WIMP) dark 
matter (DM) with DM masses $m_X \sim \mathcal{O}(1 - 100)$ keV from leptonic decays 
$M\rightarrow \ell \bar{\nu}_\ell + X$, where $M=B^\pm, D^\pm, D_s^\pm$ is a heavy meson state. 
We focus on two cases where $X$ denotes either a light pseudoscalar (axion-like), or a 
light vector state that couples to the standard model (SM) through kinetic mixing. We note that 
for a small DM mass these decays are separately sensitive to DM couplings to quarks, but not its mass.
\end{abstract}

\def\thepage{{}}
\maketitle
\def\thepage{\arabic{page}}

\section{Introduction}

There is evidence that the amount of dark matter (DM) in the Universe by far dominates that of the luminous matter. It comes  
from a variety of cosmological sources such as the rotation curves of galaxies~\cite{DMgalaxies}, 
gravitational lensing~\cite{DMlensing}, features of CMB~\cite{Komatsu:2010fb} and large scale structures~\cite{Allen:2002eu}. 
While the presence of DM is firmly established, its basic properties are still subject of a debate. If dark matter is comprised of
some fundamental particle, experimentally-measured properties, such as its relic abundance or production cross-sections
can be predicted. Experimental measurements of the abundance $\Omega_{DM} h^2 \sim 0.12$ by WMAP 
collaboration~\cite{WMAP} can be used to place constraints on the masses and interaction strengths of those DM
particles. Indeed, the relation
\begin{equation}\label{RelDen}
\Omega_{DM} h^2 \sim \langle \sigma_{ann} v_{rel}\rangle^{-1} \propto \frac{M^2}{g^4},
\end{equation}
with $M$ and $g$ being the mass and the interaction strength associated with DM
annihilation, implies that, for a weakly-interacting massive particle (WIMP) of DM, the mass scale should be
set around the electroweak scale. Yet, difficulties in understanding of small-scale gravitational 
clustering in numerical simulations with WIMPs may lead to preference being given to much lighter 
DM particles. Particularly there has been interest in studying models of light dark matter particle with 
masses of the keV range~\cite{Pospelov:2008jk,Pospelov:2007mp}. According to Eq.~(\ref{RelDen}),
the light mass of dark matter particle then implies a superweak interaction between the dark matter 
and standard model (SM) sector~\cite{SWIMP:1}. Several models with light $\cal{O}$(keV-MeV) DM particles, 
or super-WIMPs, have been proposed~\cite{Pospelov:2008jk,Pospelov:2007mp}.  

One of the main features of the super-WIMP models is that DM particles do not need to be stable against decays 
to even lighter SM particles~\cite{Pospelov:2008jk}. This implies that one does not need to impose an ad-hoc $Z_2$ 
symmetry when constructing an effective Lagrangian for DM interactions with the standard model fields, so
DM particles can be emitted and absorbed by SM particles. 
Due to their extremely small couplings to the SM particles, experimental searches for super-WIMPs must be 
performed at experiments where large statistics is available. In addition, the experiments must be able to resolve
signals with missing energy~\cite{Batell:2009yf}. Super-B factories fit this bill perfectly.

In this paper we focus on bosonic super-WIMP models~\cite{Pospelov:2008jk,Pospelov:2007mp} for dark matter 
candidates and attempt to constrain their couplings with the standard model through examining leptonic 
meson decays. The idea is quite straightforward. In the standard model the leptonic decay width of, say, a $B$-meson, 
i.e. the process $B\rightarrow \ell\bar{\nu}$, is helicity-suppressed by $(m_\ell/m_B)^2$ due to the left-handed 
nature of weak interactions~\cite{Rosner:2010ak},
\begin{equation}\label{LeptRate}
\Gamma(B\rightarrow \ell\bar{\nu}) = \frac{G_F^2}{8\pi}|V_{ub}|^2 f_B^2 m_B^3 
\frac{m_\ell^2}{m_B^2}\biggl(1-\frac{m_\ell^2}{m_B^2}\biggr)^2.
\end{equation}
Similar formulas are available for charmed meson $D^+$ and $D_s$ decays with obvious substitution of parameters.
The only non-perturbative parameter affecting Eq.~(\ref{LeptRate}), the heavy meson decay constant $f_B$, can be reliably 
estimated on the lattice~\cite{Davies:2010ip},  so the branching ratio for this process can be predicted quite reliably.

\begin{figure}
\begin{center}
\subfigure[]{\includegraphics[width=7.0cm]{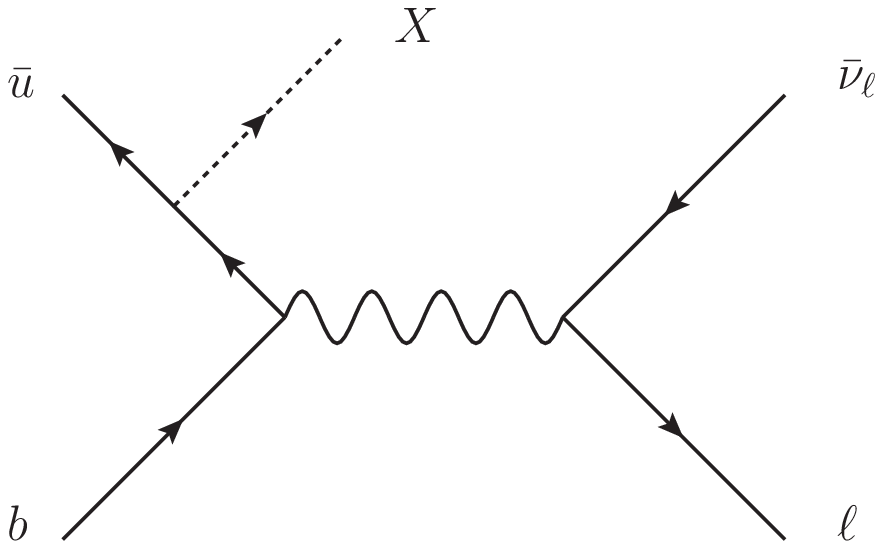}}
\subfigure[]{\includegraphics[width=7.0cm]{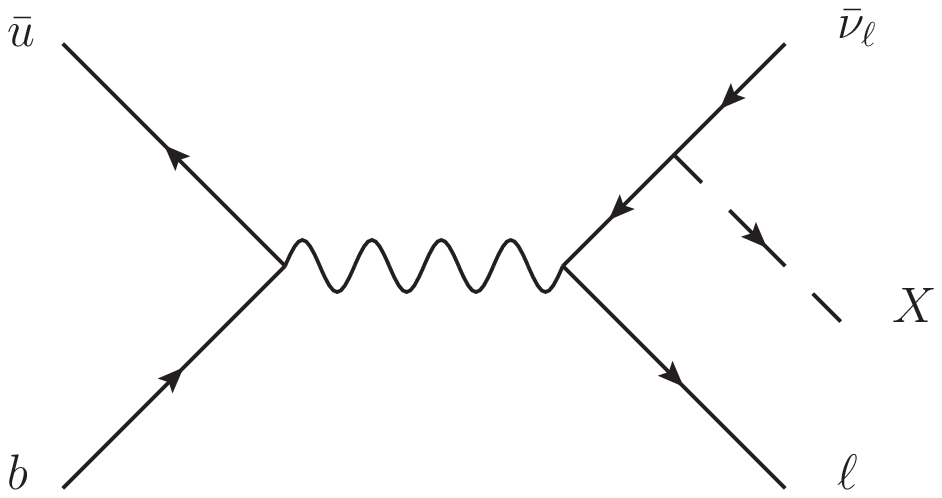}}
\subfigure[]{\includegraphics[width=7.0cm]{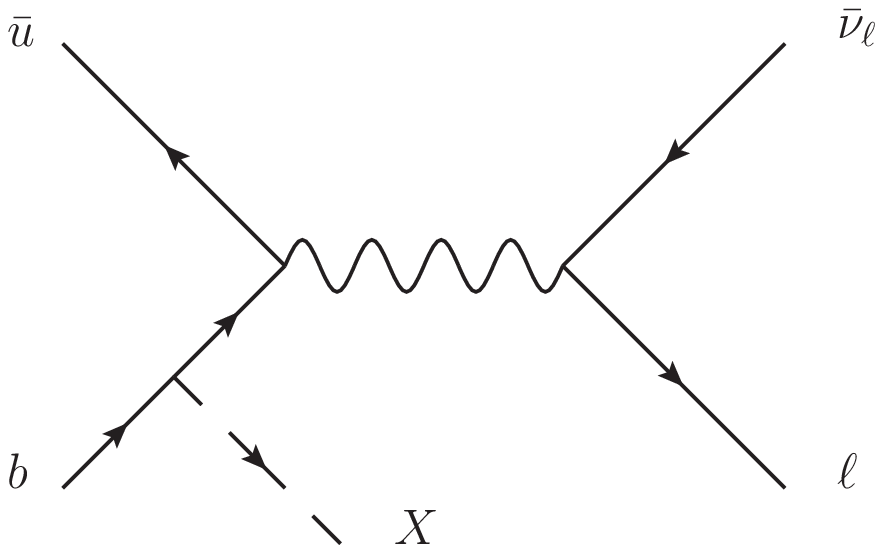}}
\subfigure[]{\includegraphics[width=7.0cm]{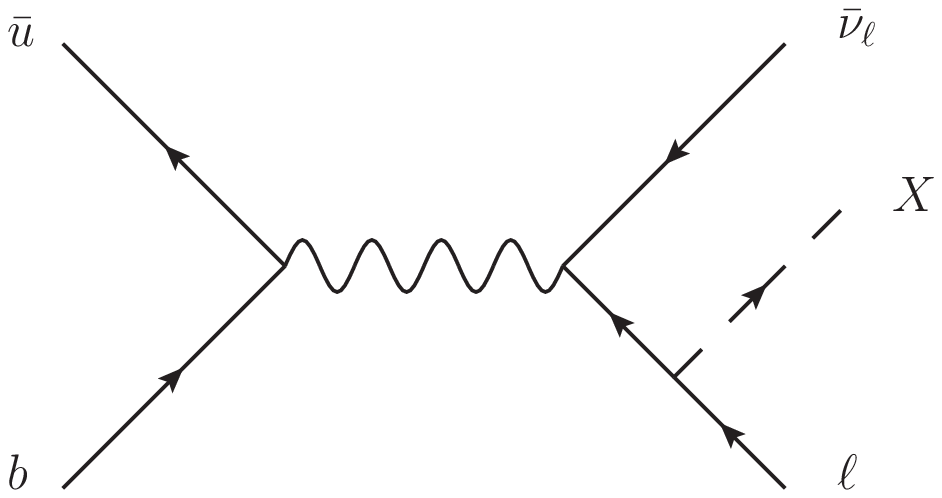}}
\end{center}
\caption{Diagrams for the super-WIMP emission in $B \to \ell \bar{\nu}_\ell X$. Similar diagrams exist for 
$D_{(s)}$ decays. Note that the graph (b) is absent for the vector light dark matter particles discussed in 
section~\ref{VectorDM}.}
\label{FIG:FeynmanDiags} 
\end{figure}

The helicity suppression arises from the necessary helicity flip on the outgoing lepton due to angular momentum conservation as 
initial state meson is spinless. The suppression can be overcome by introducing a third particle to the final state that contributes 
to total angular momentum~\cite{Burdman:1994ip} (see Fig.~\ref{FIG:FeynmanDiags}). 
If that particle is a light DM candidate, helicity suppression is traded for 
a small coupling strength of DM-SM interaction. In this case, the  charged lepton spectrum of the 3-body 
$B\rightarrow \ell \bar{\nu}_\ell + X$ (with $X$ being the DM candidate) process will be markedly different from 
the spectrum of two-body $B\rightarrow \ell \bar{\nu}_\ell$ decay. Then, the rate for the process 
$B \to \ell+\slashed E$, with $\slashed E$ being missing energy, can be used to constrain properties of light DM particles.

We shall consider two examples of super-WIMPs, the ``dark photon" spin-1 particle, and a spin-0, axion-like state.
The discussion of the vector dark matter effects is similar to a calculation of  the radiative leptonic 
decay~\cite{Burdman:1994ip}, i.e. the spin of the added DM particle brings the required unit of angular momentum.  
In the case of axion-like DM candidate, there is a derivative 
coupling to the SM allowing the pseudoscalar particle to carry orbital angular momentum and hence overcome helicity 
suppression as well. As a side note, we add that the models of new physics considered here are very different 
from the models that are usually constrained in the new physics searches with leptonic decays of heavy 
mesons~\cite{Dobrescu:2008er}.

This paper is organized as follows. In Section~\ref{AxionLike} we examine the decay width for the process 
$M\rightarrow \ell \bar{\nu}_\ell + X$ for $X=a$ being a spin-0 particle. We consider a particular two-Higgs doublet 
model, taking into account DM-Higgs mixing in Section~\ref{AxonHiggs}. In Section~\ref{VectorDM} we consider
constraints on a spin-1 super-WIMP candidate. We conclude in Section~\ref{Conclusions}.

\section{Simple Axion-Like Dark Matter}\label{AxionLike}

We consider first an ``axion-like" dark matter (ALDM) model, as suggested in \cite{Pospelov:2008jk} and study the tree-level 
interactions with the standard model fermions. The most general Lagrangian consists of a combination of dimension-five operators,
\beq
\mathcal{L}_a = -\frac{\partial_\mu a}{f_a} \bar{\psi} \gamma^\mu \gamma_5 \psi + \frac{C_\gamma}{f_a} a F_{\mu \nu} \widetilde{F}^{\mu \nu},
\eeq
where $X=a$ is the DM particle and the coupling constant $f_a$ has units of mass. Taking into account the chiral anomaly we can substitute the second term with a combination of vector and axial-vector fermionic currents,
\beq
\mathcal{L}_a = -\biggl(\frac{1}{f_a}+ \frac{4 \pi C_\gamma}{f_a \alpha}\biggr) \partial_\mu a \  \bar{\psi} \gamma^\mu \gamma_5 \psi - i m_\psi \biggl(\frac{8 \pi C_\gamma}{f_a \alpha}\biggr) a  \bar{\psi} \gamma_5 \psi.
\eeq
The Feynman diagrams that contribute to the meson decay, for  example  $B\rightarrow \ell \bar{\nu}_\ell + a$, are shown by Fig \ref{FIG:FeynmanDiags}. The amplitude for the emission of $a$ in the transition $M\rightarrow \ell \bar{\nu}_\ell + a$ can be written as
\beq
{\cal A}_{M\to \ell\bar\nu a} = {\cal A}_\ell + {\cal A}_q,
\eeq
where ${\cal A}_q$, the quark contribution, represents emission of $a$ from the quarks that build up the meson and ${\cal A}_\ell$, the leptonic contribution, describe emission of $a$ from the final state leptons. 

Let's consider the lepton amplitude first. Here we can parameterize the axial matrix elements contained in the amplitude in terms of the decay constant $f_B$ such as
\beqa
\langle 0| \bar{u} \gamma^\mu \gamma_5 b| B\rangle = i f_B P_B^\mu, 
\eeqa
If the mass of the axion-like DM particle is small ($m_a\rightarrow0$), the leptonic contribution simplifies to 
\beq
\mathcal{A}_{\ell} = i \sqrt{2}  G_F V_{ub} \frac{f_B}{f_a}  m_{\ell} 
\biggl( \frac{m_\ell}{2 k \cdot p_\ell} [\bar{u}_\ell \not k (1-\gamma_5) v_\nu] - [\bar{u}_\ell (1-\gamma_5) v_\nu] \biggr).
\eeq
Here $k$ is the DM momentum. Clearly, this contribution is proportional to the lepton mass and can, in principle, 
be neglected in what follows. The contribution to the decay amplitude from the DM emission from the quark current is
\beq
\mathcal{A}_{q} = i \langle 0|\bar{u}\Gamma^\mu b| B\rangle [\bar{u}_\ell \gamma_\mu (1-\gamma_5) v_\nu]
\eeq
where the current $\bar{u}\Gamma^\mu b$ is obtained from the diagrams in Figure~\ref{FIG:FeynmanDiags} (a) and (c),
\beq 
\Gamma^\mu =  \frac{G_F}{\sqrt{2}f_a} V_{ub}
\biggl[ \frac{(\not k \gamma_5)(\not k  - \not{p_u} + m_u)\gamma^\mu(1-\gamma_5)}{m_a^2 - 2p_u\cdot k} + 
\frac{\gamma^\mu(1-\gamma_5)(\not{p_b}  - \not k + m_b)(\not k \gamma_5)}{m_a^2 - 2p_b\cdot k}\biggr] .
\eeq
Since the meson is a bound state of quarks we must use a model to describe the effective quark-antiquark distribution. 
We choose to follow Refs.~\cite{Szczepaniak:1990dt} and \cite{Lepage:1980fj}, where the wave function for a
ground state meson $M$ can be written in the form
\beq
\psi_M = \frac{I_c}{\sqrt{6}} \phi_M(x) \gamma_5 (\not{P_M} + M_M g_M(x)).
\eeq
Here $I_c$ is the identity in color space and $x$ is the momentum fraction carried by one of the quarks. For a heavy 
meson $H$ it would be convenient to assign $x$ as a momentum fraction carried by the heavy quark.
Also, for a heavy meson, $g_H \sim 1$, and in the case of a light meson $g_L = 0$.
For the distribution amplitudes of a heavy or light meson we use
\beqa
\phi_L \ &\sim& \ x(1-x),\\
\phi_H \ &\sim& \ \left[\frac{(m^2/M_H^2)}{1-x} + \frac{1}{x} - 1\right]^{-2},
\eeqa
where $m$ is the mass of the light quark and the meson decay constant is related to the normalization of the 
distribution amplitude,
\beq
\int^1_0 \phi_M(x)dx = \frac{f_M}{2\sqrt{6}}.
\eeq
The matrix element can then be calculated by integrating over the momentum fraction \cite{Lepage:1980fj}
\beq
 \langle 0|\bar q \Gamma^\mu Q | M\rangle = \int^1_0 dx \ \mbox{Tr}\left[\Gamma^\mu \psi_M\right].
 \eeq
Neglecting the mass of the axion-like DM particle, the decay amplitude in the $B^\pm$ case simplifies to 
\beqa
\label{eqn_mesonamp}
\mathcal{A}_{q} =  i \frac{\sqrt{3} G_F V_{ub} M_B}{f_a(k\cdot P_B)}
\left(M_B \Phi_1^B - m_b \Phi_0^B\right)\left[\bar{\ell}\not k (1-\gamma_5)\nu\right],
\eeqa
where $m_b$ is the mass of the $b$-quark (or, in general, a down-type quark in the decay), and we defined
\beq
\Phi_n^M = \int^1_0 \frac{\phi_M(x)}{x(1-x)} x^n dx
\eeq
The total decay width is, then,
\beqa\label{BtoAxion}
\Gamma_{B\rightarrow a \ell \nu_\ell } &=&  \frac{G_F^2 f_B^2 |V_{ub}|^2 M_B^5}{64 \pi^3 f_a^2}
\left[ \frac{1}{6} (2\rho^2 + 3 \rho^4 + 12 \rho^4 \log \rho - 6\rho^6 + \rho^8) \right. 
\nonumber \\
 &+& \left. g_B^2 \ \Phi(m_b, M_B)^2
 (1- 6\rho^2 - 12 \rho^4 \log \rho + 3 \rho^4 + 2\rho^6) 
 \right],
\eeqa
where $\rho \equiv m_{\ell}/m_B$. Also,
\beq
\Phi(m_b, M_B)   = \frac{m_b\Phi_0 - M_B\Phi_1}{f_B M_B}.
\eeq
Note that $\Phi(m_b, M_B) \propto 1/m$, which is consistent with spin-flipping transition in a quark model, 
which would explain why this part of the decay rate is not proportional to $m_\ell$.
Similar results for other heavy mesons, like $D^+$ and $D^+_s$ are obtained by the obvious substitution
of relevant parameters, such as masses, decay constants and CKM matrix elements. 

Experimentally, the leptonic decays of heavy mesons are best studied at the $e^+e^-$ flavor factories 
where a pair of $M^+M^-$ heavy mesons are created. The study is usually done by fully 
reconstructing one of the heavy mesons and then by finding a candidate lepton track of opposite sign to 
the tagged meson. The kinematical constraints on the lepton are then used to identify the decays 
with missing energy as leptonic decay. 

\begin{figure}
\begin{center}
\subfigure[~~$B^{\pm} \rightarrow \ell \bar{\nu}_\ell a$]{\includegraphics[width=6.5cm]{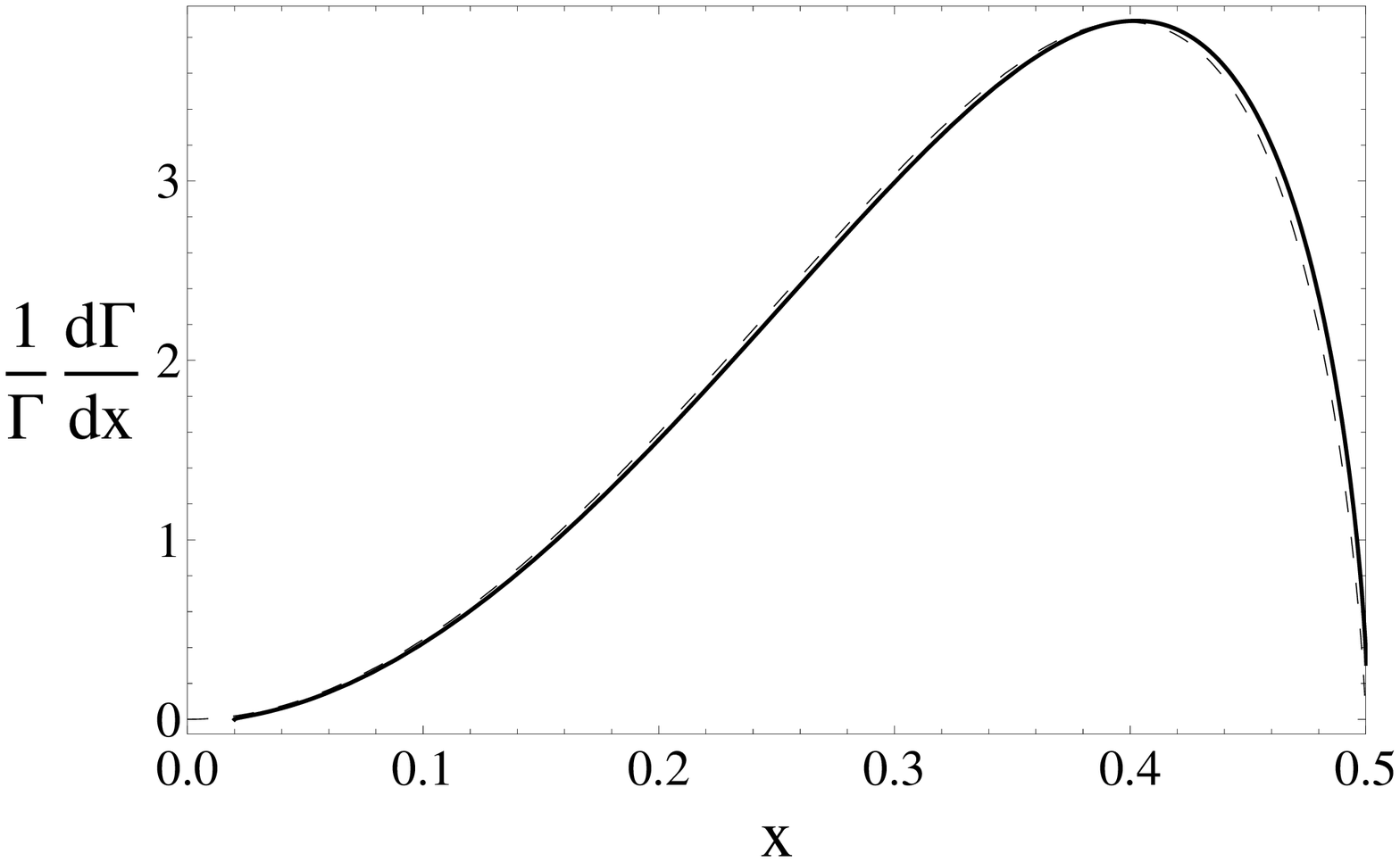}}
\subfigure[~~$D^{\pm} \rightarrow \ell \bar{\nu}_\ell a$]{\includegraphics[width=6.5cm]{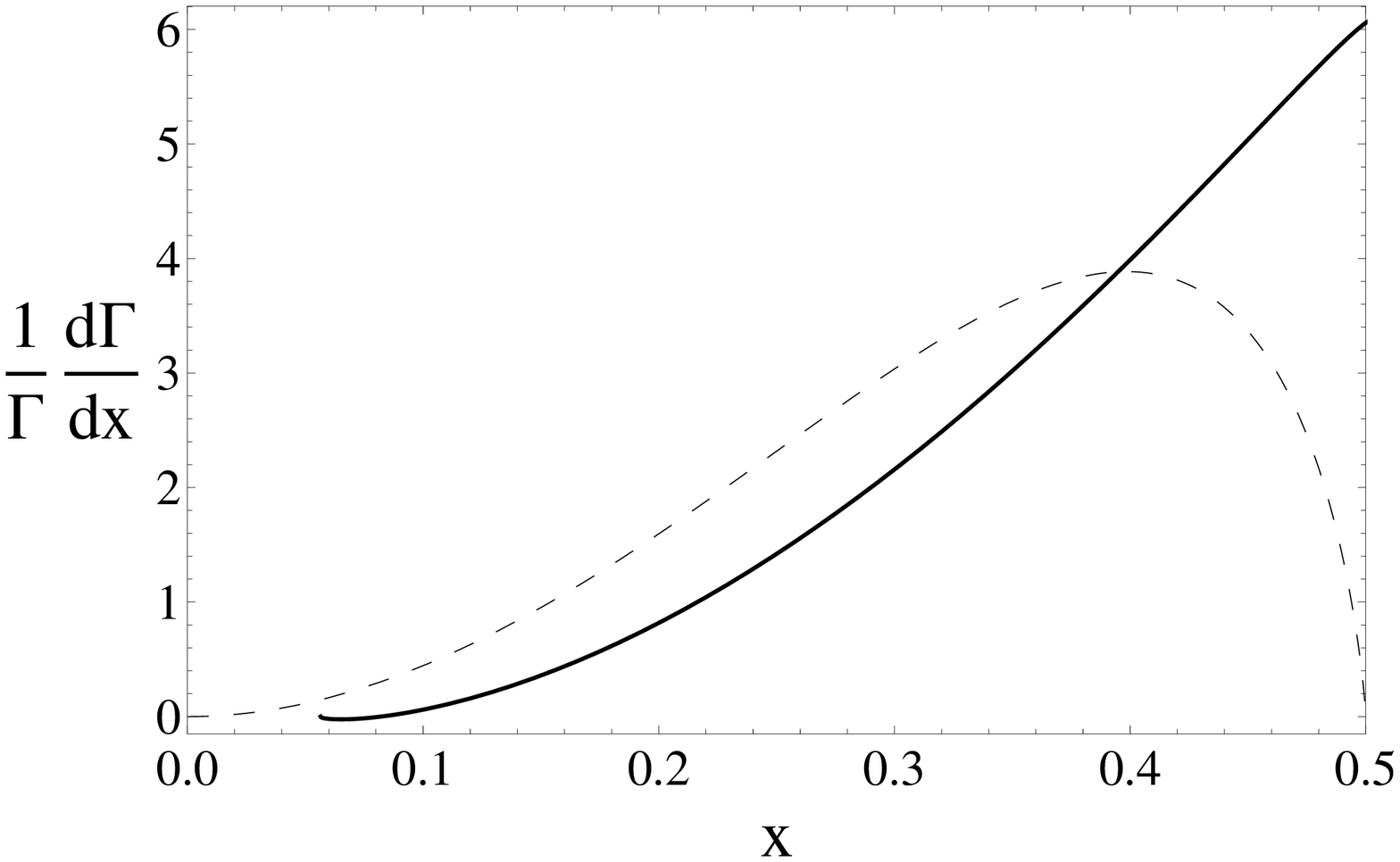}}
\subfigure[~~$D_s^{\pm} \rightarrow \ell \bar{\nu}_\ell a$]{\includegraphics[width=6.5cm]{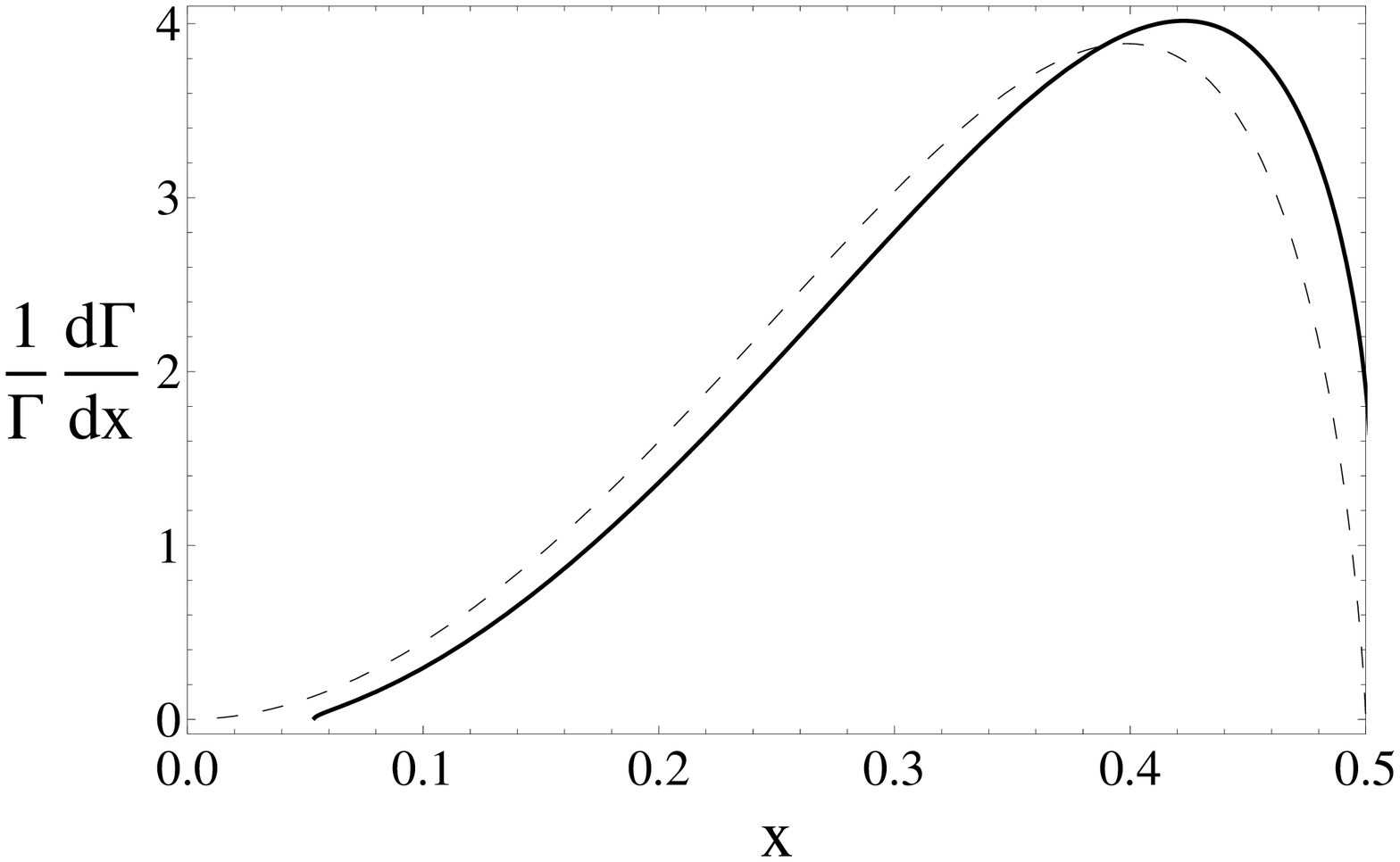}}
\end{center}
\caption{Normalized electron (dashed) and muon (solid) energy distributions for the 
heavy ($B^{\pm},D^{\pm},D_s^{\pm}$) meson decay channels. Here $m_a=0$ and $x=E_\ell/m_B$.}
\label{FIG:ALDM_Lept_Dist} 
\end{figure}

In the future super-B factories, special studies of the lepton spectrum in $M \to \ell +$missing energy can be 
done using this technique to constrain the DM parameters from Eq.~(\ref{BtoAxion}). The 
lepton energy distributions, which are expected to quite different for the three-body decays 
$B^{-}\rightarrow a \ell^{-} \bar{\nu}_\ell$ are shown (normalized) in Fig.~\ref{FIG:ALDM_Lept_Dist} for 
each lepton decay process. However, we can 
put some constraints on the DM coupling parameters using the currently available data on $M \to \ell\bar\nu_\ell$.
The experimental procedure outlined above implies that what is experimentally detected is the combination,
\beqa\label{MasterFormula}
\Gamma_{\mbox{\tiny exp}}(M \to  \ell\bar\nu_\ell) \ &=& \
\Gamma_{\mbox{\tiny SM}}(M \to  \ell\bar\nu_\ell) + \int_{E<E_0} dE_a \frac{d \Gamma(M \to  a \ell\bar\nu_\ell)}{d E_a} 
\nonumber \\ 
 &=&  \ \Gamma_{\mbox{\tiny SM}}(M \to  \ell\bar\nu_\ell) \
\left[1 + R_a(E_0) \right],
\eeqa
where $E_0$ is the energy cutoff that is specific for each experiment. Equivalently, cutoff in $q^2$ can also be used. 
In the above formula we defined
\beq
R_a (E_0) = \frac{1}{\Gamma_{\mbox{\tiny SM}}(M \to  \ell\bar\nu_\ell)} 
\int_{E<E_0} dE_a \frac{d \Gamma(M \to  a \ell\bar\nu_\ell)}{d E_a} .
\eeq
Our bounds on the DM couplings from different decay modes are reported in Table~\ref{FIG:BRALDM} for the
cutoff values of $E_0 = 100$~MeV. 
\begin{table}
\begin{center}
\footnotesize
\begin{tabular}{|c||c|c|c||c|c|c|}
\hline\hline
\begin{tabular}{c} Channel\\(Seen)\end{tabular}   &    \begin{tabular}{c} Experiment\\(Maximum)\end{tabular}  & \begin{tabular}{c} Standard\\Model\end{tabular} & \begin{tabular}{c} $ f_a^{2} R_a(E_0)$ \\ $E_0= 100~\mbox{MeV}$\end{tabular} & \begin{tabular}{c} $R_{\gamma_s}(E_0^\prime)$\\  $E_0^\prime= 50~\mbox{MeV}$   \end{tabular} &  \begin{tabular}{c} $R_{\gamma_s}(E_0^\prime)$ \\ $E_0^\prime= 100~\mbox{MeV}$   \end{tabular} &  \begin{tabular}{c} $R_{\gamma_s}(E_0^\prime)$ \\ $E_0^\prime= 300~\mbox{MeV}$  \end{tabular} \\
\hline\hline
$\mathcal{B}\left(B^{\pm}\rightarrow \tau^{\pm} \bar{\nu}_\tau \right)$ & $1.7\times 10^{-4}$&~$7.9\times 10^{-5}$&$1.6\times 10^2$&$4.9\times 10^{-5}$&$ 1.9\times 10^{-4} $&$ 1.9 \times 10^{-3} $  \\
$\mathcal{B}\left(D^{\pm}\rightarrow \mu^{\pm} \bar{\nu}_\mu \right)$ &
$3.8\times 10^{-4}$&$3.6\times 10^{-4}$&$3.1\times 10^3$&$ 4.0\times 10^{-3} $&~$ 1.8 \times 10^{-2} $&$ 1.7 \times 10^{-2} $  \\
$\mathcal{B}\left(D_s^{\pm}\rightarrow \mu^{\pm} \bar{\nu}_\mu \right)$ & 
$5.9\times 10^{-3}$&$5.3\times 10^{-3}$&$4.6\times 10^2$&$2.0\times 10^{-4} $&~$ 7.8 \times 10^{-4} $&$ 6.0 \times 10^{-3}$ \\
$\mathcal{B}\left(D_s^{\pm}\rightarrow \tau^{\pm} \bar{\nu}_\tau \right)$ &
$5.4\times 10^{-2}$&$5.1\times 10^{-2}$&$6.5\times 10^0$&$2.1 \times 10^{-5}$&~$ 8.0 \times 10^{-5} $&$ 6.2 \times 10^{-4} $  \\
\hline\hline Channel (Unseen)   &&&&&&\\
\hline\hline
$\mathcal{B}\left(B^{\pm}\rightarrow e^{\pm} \bar{\nu}_e \right)$ & $~< 1.9\times 10^{-6}$&~$8.3\times 10^{-12}$&$6.6\times 10^7$&$ 4.6 \times 10^2 $&$ 1.8 \times 10^3 $&$ 1.6 \times 10^4$  \\
$\mathcal{B}\left(B^{\pm}\rightarrow \mu^{\pm} \bar{\nu}_\mu \right)$ &
$~< 1.0 \times 10^{-6}$&$3.5\times 10^{-7}$&$1.8 \times 10^3 $&$1.1 \times 10^{-2} $&$ 4.3  \times 10^{-2} $&$ 3.6  \times 10^{-1} $  \\
$\mathcal{B}\left(D^{\pm}\rightarrow e^{\pm} \bar{\nu}_e \right)$ & 
$~< 8.8\times 10^{-6}$&$8.5\times 10^{-9}$&$3.1\times 10^6$&$1.9\times 10^{2} $&$ 7.6 \times 10^{2} $&$ 7.1 \times 10^{3} $ \\
$\mathcal{B}\left(D^{\pm}\rightarrow \tau^{\pm} \bar{\nu}_\tau \right)$ &
$~< 1.2\times 10^{-3}$&$9.7\times 10^{-4}$&$1.0\times 10^{1}$&$1.7  \times 10^{-3} $&$ 7.7  \times 10^{-3} $&$ 6.2  \times 10^{-2} $  \\
$\mathcal{B}\left(D_s^{\pm}\rightarrow e^{\pm} \bar{\nu}_e \right)$ &
$~< 1.2\times 10^{-4}$&$1.2\times 10^{-7}$&$9.8	\times 10^6$&$8.6  \times 10^{0} $&$ 3.3  \times 10^{1} $&$ 2.6 \times 10^2 $  \\
\hline\hline
\end{tabular}
\normalsize
\end{center}
\caption{Constraints on $f_a$ from various decays. The last three columns represent possible soft photon
pollution of $M \to \ell\bar\nu_\ell$ decays for three different values of photon energy cutoff.}
\label{FIG:BRALDM} 
\end{table}
Note that similar expressions for the leptonic decays of the {\it light} mesons, such as $\pi \to a\ell\bar\nu$ and 
$K \to a\ell\bar\nu$ come out to be proportional to the mass of the final state lepton. This is due to the fact that in the 
light meson decay the term proportional to $g$ vanishes. Thus, those decays do not 
offer the same relative enhancement of the three-body decays due to removal of the helicity suppression in the 
two-body channel. It is interesting to note that the same is also true for the 
heavy mesons if a naive Non-Relativistic Constituent Quark Model (NRCQM), similar to the one used in 
Refs.~\cite{Chang:1997re,Lu:2002mn} is employed. We checked that a simple replacement 
\beqa
p_b = \frac{m_b}{m_B} P_B,\qquad \qquad p_u = \frac{m_u}{m_B} P_B
\eeqa
advocated in \cite{Chang:1997re,Lu:2002mn} is equivalent to use of a symmetric (with respect to the momentum fraction carried 
by the heavy quark) distribution amplitude, which is not true in general.

Currently, the SM predictions for the $B^{-}\rightarrow \ell^{-} \bar{\nu}_\ell$ decay for $\ell = \mu, e$ are significantly 
smaller than the available experimental upper bounds~\cite{Asner:2010qj,PDG2010:1}, which is due to the smallness of $V_{ub}$
and the helicity suppression of this process. Thus, even in the standard model, there is a possibility that some of the processes
$B^{-}\rightarrow  \gamma_s \ell^{-} \bar{\nu}_\ell$, with $\gamma_s$ being the soft photon , are missed by the experimental
detector. Such photons would affect the bounds on the DM couplings reported in Table~\ref{FIG:BRALDM}. 

The issue of the soft photon ``contamination" of $B^{-}\rightarrow \ell^{-} \bar{\nu}_\ell$ is non-trivial if model-independent 
estimates of the contributions are required (for the most recent studies, see \cite{Becirevic:2009aq}). In order to take those
into account, the formula in Eq.~(\ref{MasterFormula}) should be modified to
\beq\label{MasterFormula2}
\Gamma_{\mbox{\tiny exp}}(M \to  \ell\bar\nu_\ell) \ = \
\Gamma_{\mbox{\tiny SM}}(M \to  \ell\bar\nu_\ell) \
\left[1 + R_a(E_0) + R_{\gamma_s} (E_0^\prime) \right].
\eeq
In general, the experimental soft photon cutoff $E_0^\prime$ could be different from the DM emission cutoff $E_0$. 
Since we are only interested in the upper bounds on the DM couplings, this issue is not very relevant here, as the 
amplitudes with soft photons do not interfere with the amplitudes with DM emission. Nevertheless, 
for the purpose of completeness, we evaluated the 
possible impact of undetected soft photons using NRCQM as seen in \cite{Chang:1997re,Lu:2002mn}. The results are 
presented in Table~\ref{FIG:BRALDM} for different values of cutoff on the photon's energy. We present the NRCQM
mass parameters in Table~ \ref{Min:datavalues} with the decay constants calculated in  \cite{Lucha:2010dh}.
The relevant plots for $D$ ($D_s$) decays can be obtained upon substitution $M_B \to M_{D(D_s)}$, 
$f_B \to f_{D(D_s)}$, and $V_{ub} \to V_{cd(cs)}$. Note that there is no CKM suppression for $D_s$ decays.
In order to bound $f_a$ we use the experimentally seen transitions $B \to \tau \bar\nu$, $D_{(s)} \to \mu \bar\nu$, 
and $D_s \to \tau \bar\nu$. We note that the soft photon ``contamination" can be quite large, up to 
$10\%$ of the standard model prediction for the two body decay.
\begin{table}
\begin{center}
\small
\begin{tabular}{|c|c|}
\hline\hline
~Quark~ & ~Constituent Mass~ \\
\hline\hline
$m_u$ & $335.5$ MeV   \\
$m_d$ & $339.5$ MeV   \\ 
$m_s$ & $486 $ MeV   \\ 
$m_c$ & $1550$ MeV     \\
$m_b$ & $4730$ MeV     \\
\hline\hline
\end{tabular}
\normalsize
\end{center}
\caption{Constituent quark masses \cite{Scadron:2006dy} used in calculations. }
\label{Min:datavalues} 
\end{table}
The resulting fits on $f_a$ can be found in Table~\ref{FIG:ALDM_fa_fits}. As one can see, the best constraint 
comes from the $D^{\pm} \rightarrow \mu^{\pm} \bar{\nu}_\mu $ decay where experimental and theoretical 
branching ratios are in close agreement.  
\begin{table}
\begin{center}
\footnotesize
\begin{tabular}{|c|c|}
\hline\hline
Channel & $f_a, MeV$\\
\hline
\hline
~~$\mathcal{B}\left(B^{\pm}\rightarrow 
\tau^{\pm} \bar{\nu}_\tau \right)$ ~~& 12 \\
~~$\mathcal{B}\left(D^{\pm}\rightarrow 
\mu^{\pm} \bar{\nu}_\mu \right)$ ~~& 236 \\
~~$\mathcal{B}\left(D_s^{\pm}\rightarrow 
\mu^{\pm} \bar{\nu}_\mu \right)$ ~~& 62 \\
~~$\mathcal{B}\left(D_s^{\pm}\rightarrow 
\tau^{\pm} \bar{\nu}_\tau \right)$ ~~& 11 \\
\hline\hline
\end{tabular}
\normalsize
\end{center}
\caption{Constraint on $f_a$ using the various seen decay channels. }
\label{FIG:ALDM_fa_fits} 
\end{table} 
%

 \section{Axion-like Dark Matter in a Type II Two Higgs Doublet Models}\label{AxonHiggs}
 
A generic axion-like DM considered in the previous section was an example of a simple augmentation of the 
standard model by an axion-like dark matter particle. A somewhat different picture can emerge if those 
particles are embedded in more elaborate beyond the standard model (BSM) scenarios. For example, in 
models of heavy dark matter of the ``axion portal"-type~\cite{Nomura:2008ru}, spontaneous breaking of the 
Peccei-Quinn (PQ) symmetry leads to an axion-like particle that  can mix with the CP-odd Higgs $A^0$ of a 
two Higgs Doublet model (2HDM). For the sufficiently small values of its mass this state itself can play a role 
of the light DM particle.  The decays under consideration can be derived from the $B\rightarrow \ell \nu A^0$ 
amplitude. An interesting feature of this model is the dependence of the light DM coupling upon the quark mass.
This means that the decay rate would be dominated by the contributions enhanced by the heavy quark mass. 
This would also mean that the astrophysical constraints on the axion-like DM parameters might not probe 
all of the parameter space in this model. 
  
In a concrete model~\cite{Nomura:2008ru}, the PQ symmetry $U(1)_{PQ}$ is broken by a large vacuum 
expectation value $\langle S \rangle \equiv f_a \gg v_{EW}$ of a complex scalar singlet $\Phi$. As 
in \cite{Freytsis:2009ct},  we shall work in an interaction basis so that the axion state appears in $\Phi$ as  
  \beq
         \Phi = f_a \exp \left[\frac{i a}{\sqrt{2} f_a}\right]
  \eeq
and $A^0$ appears in the Higgs doublets in the form
  \beqa
         \Phi_u= \left(\begin{array}{c}  v_u \exp \biggl[\frac{i\cot \beta}{\sqrt{2}v_{EW}}A^0 \biggr] \\0\end{array}\right), \qquad
            \Phi_d= \left(\begin{array}{c}  0\\v_d \exp \biggl[\frac{i\tan \beta}{\sqrt{2}v_{EW}}A^0 \biggr] \end{array}\right),
    \eeqa
where we suppress the charged and CP-even Higgses for simplicity and define $\tan \beta = v_u/v_d$ and 
$v_{EW} = \sqrt{v_u^2 + v_d^2} \equiv \frac{m_W}{g}$. We choose the operator that communicates PQ charge 
to the standard model  to be of the form\footnote{This is the case of the so-called Dine-Fischler-Srednicki-Zhitnitsky 
(DFSZ) axion, although other forms of the interaction term with other powers of the scalar field $\Phi$ are 
possible~\cite{Freytsis:2009ct}.}
\beq
     \mathcal{L} = \lambda \Phi^2 \Phi_u \Phi_d + h.c.
  \eeq
This term contains the mass terms and, upon diagonalizing, the physical states in this basis are given by~\cite{Freytsis:2009ct}
\beqa
a_p \ &=& \ a \cos\theta - A^0 \sin\theta \\
A^0_p \ &=& \ a \sin\theta + A^0 \cos\theta
\eeqa 
where $\tan\theta = (v_{EW}/f_a)\sin{2\beta}$. Here $a_p$ denotes the ``physical" axion-like state. 
Thus, the amplitude for $B\rightarrow \ell\nu a_p$ can be derived from 
\beq
    \mathcal{M}(B\rightarrow \ell\nu a_p) = - \sin\theta \mathcal{M}(B\rightarrow \ell\nu A^0) + 
    \cos\theta \ \mathcal{M}(B\rightarrow \ell\nu a)
\eeq 
In a type II 2HDM \cite{Freytsis:2009ct,Branco:2011iw}, the relevant Yukawa interactions of the CP-odd Higgs with fermions are given by
\beq
  \mathcal{L}_{A^0 f\bar{f}} = \frac{i g \tan\beta}{2m_W} m_d\bar{d} \gamma_5 d A^0 + \frac{ig\cot\beta}{2m_W} m_u\bar{u} \gamma_5 u A^0
\eeq
where $d=\{d,s,b\}$ refers to the down type quarks and $u=\{u,c,t\}$ refers to the up type quarks. 
The interaction with leptons are the same as above with $d\rightarrow\ell$ and $u\rightarrow\nu$.
  
In the axion portal scenario the axion mass is predicted to lie within a specific range of $360 < m_a \leq 800$  MeV to 
explain the galactic positron excess \cite{Nomura:2008ru}. Using the
 quark model introduced in the previous section we obtain the decay width
\beqa
\Gamma\left(B\rightarrow \ell \nu_\ell a_{p}\right) \
&=& \ \frac{ G_F^2 |V_{ub}|^2  m_B^3}{256 \pi ^3  \left(f_a^2+ v_{EW}^2\sin^2{2 \beta} \right)} 
\nonumber \\
&\times&
\left[\cos 2 \beta \left(m_u \Phi^B_1 + m_b (\Phi^B_0-\Phi^B_1)\right) + 
5 \left[ m_b (\Phi^B_1-\Phi^B_0) + m_u \Phi^B_1 \right] \right]^2\\
&\times& 
\Biggl[12 x_a ^4 \log (x_a )-4 x_a ^6+3 x_a ^4+(\rho -1)^4 (4 (\rho -2) \rho +1)-12 (\rho -1)^4 \log (1-\rho )\Biggr]
\qquad
\nonumber
\eeqa
Here we defined  $x_a = m_a/m_B$, and $\rho = m_\ell/m_B$. If we assume $f_a \gg v_{EW} \sin{2 \beta}$ we can then 
provide bounds on $f_a$ as seen in Table~\ref{TAB:CPODD_fa_fits}.
\begin{table}
\begin{center}
\footnotesize
\begin{tabular}{|c|c|c|c|c|}
\hline\hline
 &$f_a (MeV)$&$f_a (MeV)$& $f_a (MeV)$&$f_a (MeV)$\\
\hline
Channel & $\tan{\beta}=1$ & $\tan{\beta}=5$ & $\tan{\beta}=10$ & $\tan{\beta}=20$\\
\hline
\hline
$\mathcal{B}\left(B^{\pm}\rightarrow 
\tau^{\pm} \bar{\nu}_\tau \right)$ & 70&340&357&361  \\
$\mathcal{B}\left(D^{\pm}\rightarrow 
\mu^{\pm} \bar{\nu}_\mu \right)$ &  416&2874&3078&3131 \\
$\mathcal{B}\left(D_s^{\pm}\rightarrow 
\mu^{\pm} \bar{\nu}_\mu \right)$ & 532&1380&1499&1529  \\
\hline\hline
\end{tabular}
\normalsize
\end{center}
\caption{Constraint on $f_a$ using the observed decays for various $\tan{\beta}$s.}
\label{TAB:CPODD_fa_fits} 
\end{table} 
Just like in the previous section, the results for other decays, such as $D_{(s)} \to \ell \bar\nu_\ell$, 
can be obtained by the trivial substitution of masses and decay constants.

\section{Light Vector Dark Matter}\label{VectorDM}

Another possibility for a super-WIMP particle is a light (keV-range) vector dark matter boson (LVDM) coupled 
to the SM solely through kinetic mixing with the hypercharge field strength~\cite{Pospelov:2008jk}. This can be done
consistently by postulating an additional $U(1)_V$ symmetry. The relevant terms in the Lagrangian are
\beqa\label{VecDMLagr}
\mathcal{L} =  -\frac{1}{4} F_{\mu \nu} F^{\mu \nu} -\frac{1}{4} V_{\mu \nu} V^{\mu \nu} - \frac{\kappa}{2}V_{\mu \nu} F^{\mu \nu} + \frac{m_V^2}{2} V_\mu V^\mu + \mathcal{L}_{h'},
\eeqa
where $\mathcal{L}_{h'}$ contains terms with, say, the Higgs field which breaks the $U(1)_V$ symmetry,
$\kappa$ parameterizes the strength of kinetic mixing, and, for simplicity, we directly work with the photon
field $A_\mu$. In this Lagrangian only the photon $A_\mu$ fields (conventionally) couple to the SM fermion 
currents. 

It is convenient to rotate out the kinetic mixing term in Eq.~(\ref{VecDMLagr}) with field redefinitions
\beqa
A\rightarrow A' - \frac{\kappa}{\sqrt{1-\kappa^2}} V', &\qquad\qquad&
V\rightarrow \frac{1}{\sqrt{1-\kappa^2}} V'.
\eeqa
The mass $m_V$ will now be redefined as $m_V \rightarrow \frac{m_V}{\sqrt{1-\kappa^2}}$. Also, both $A_\mu^\prime$ and
$V_\mu^\prime$ now couple to the SM fermion currents via
\beq
\mathcal{L}_f = -e Q_f A'_{\mu} \bar{\psi}_f \gamma^\mu \psi_f - \frac{\kappa e Q_f}{\sqrt{1-\kappa^2}} V'_\mu \bar{\psi}_f \gamma^\mu \psi_f,
\eeq
where $Q_f$ is the charge of the interacting fermion thus introducing our new vector boson's coupling to the SM 
fermions. Calculations can be now carried out with the approximate modified charge coupling for $\kappa \ll 1$,
\beq
\frac{\kappa e}{\sqrt{1-\kappa^2}} \approx \kappa e.
\eeq
As we can see, in this case the coupling of the physical photon did not change much compared to the original field $A_\mu$,
while the DM field $V_\mu^\prime$ acquired small gauge coupling $\kappa e$. It is now trivial to calculate the process 
$B\rightarrow \ell \bar{\nu} V_{DM}$, as it can be done similarly to the case of the soft photon emission in Sect.~\ref{AxionLike}.
Employing the gauge condition $\epsilon \cdot k = 0$ for the DM fields, the amplitudes become in the limit
$m_V \rightarrow 0$
\beqa
\mathcal{A}_q &=&i\frac{G_F V_{ub} \kappa e  \epsilon^{* \alpha} }{6 k \cdot p_B}
\left[A^{\mu}_{\alpha} \bar{\ell} \gamma_\mu (1-\gamma_5) \nu_{\ell}
+ B  \bar{\ell} \gamma_\alpha (1-\gamma_5) \nu_{\ell}
+ C_{\alpha} \bar{\ell} (1-\gamma_5) \nu_{\ell},
 + D^{\mu} \bar{\ell} \sigma_{\mu \alpha} (1+\gamma_5) \nu_{\ell}
 \right]
\eeqa
with the coefficients
\beqa
A_{\alpha}^{\mu} \quad&=&\quad  \left[3 \sqrt{2} f_B - 2 \sqrt{3} (\Phi^B_0+\Phi^B_1)\right]k^{\mu} q^{\alpha} 
- 2 \sqrt{3} (\Phi^B_0-3\Phi^B_1) i \epsilon^{ \mu \alpha \sigma \rho} k_{\sigma} q_{\rho},\\
B \quad &=& \quad -\left[3 \sqrt{2} f_B - 2 \sqrt{3} (\Phi^B_0+\Phi^B_1)\right] (k\cdot q)- \frac{3}{\sqrt{2}} f_B m_B^2 
\nonumber \\
&&  \quad - \ 2 \sqrt{3} g m_B \left[m_2 (\phi_0-3 \phi_1) +2 m_B \phi_1\right],
\\
C_{\alpha} \quad&=&\quad 3 \sqrt{2} f_B m_{\ell} \frac{q^{\alpha} k\cdot p_{\ell} - p_{\ell}^{\alpha} k\cdot q}{k\cdot p_{\ell}},
\\
D^{\mu}\quad&=&\quad -3 \sqrt{2} i f_B m_{\ell} \frac{k\cdot q}{k \cdot p_{\ell}}  k^{\mu},
\eeqa
and  $q = p_{\ell} + p_{\nu}$. 
\begin{figure}
\begin{center}
\subfigure[$~B^{\pm} \rightarrow \ell \bar{\nu}_\ell V_{DM}$]{\includegraphics[width=7.5cm]{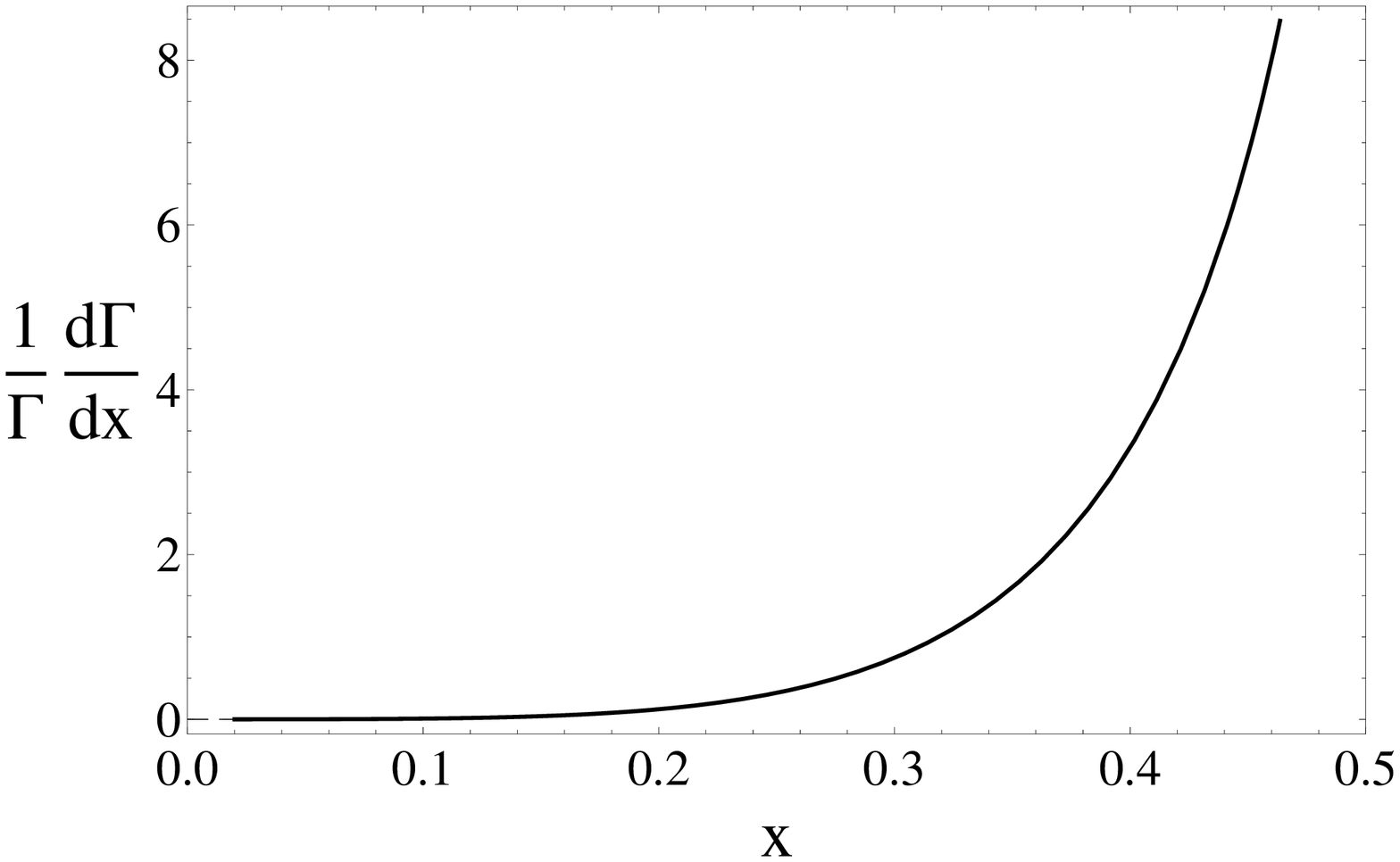}}
\subfigure[$~D^{\pm} \rightarrow \ell \bar{\nu}_\ell V_{DM}$]{\includegraphics[width=7.5cm]{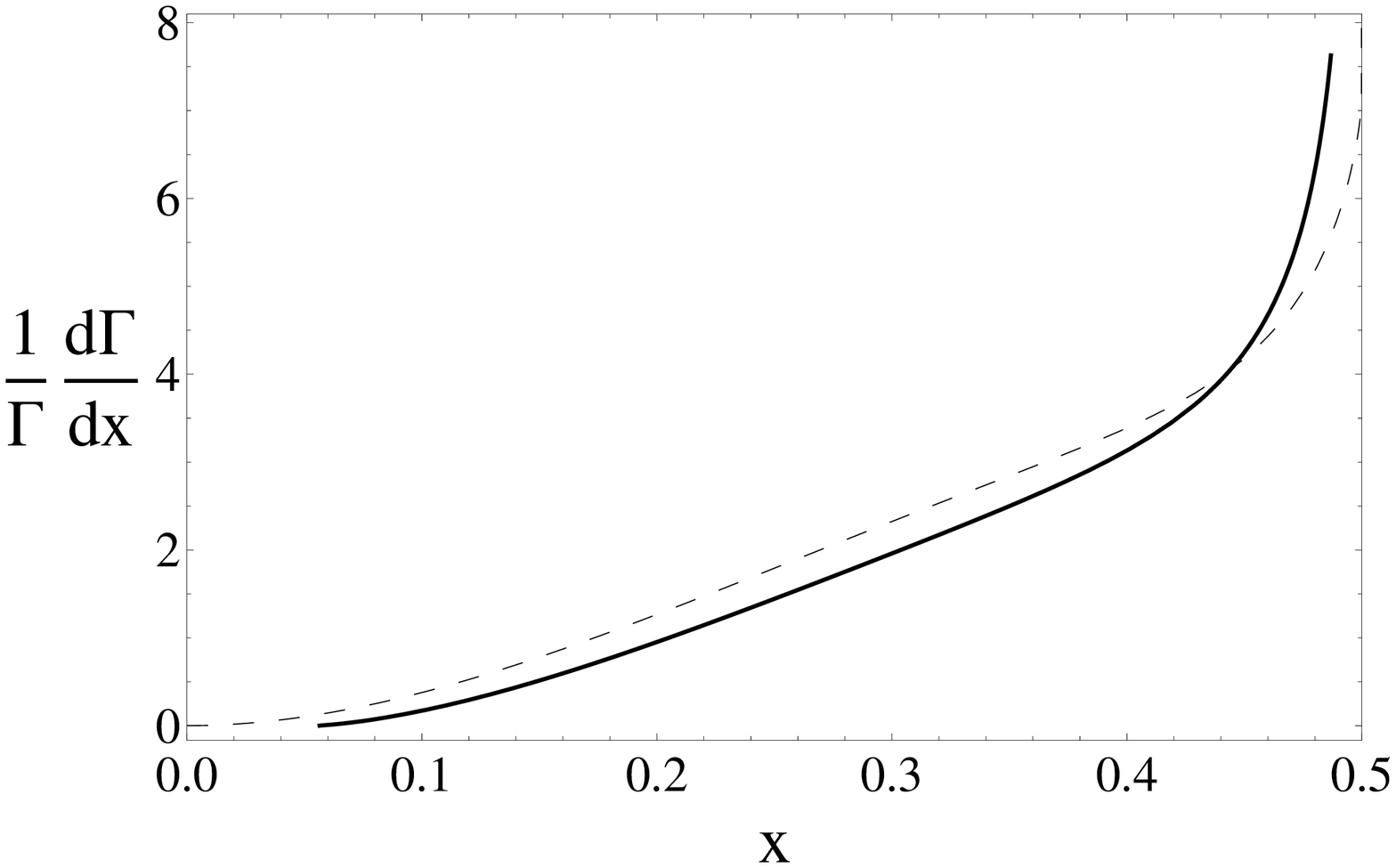}}
\subfigure[$~D_s^{\pm} \rightarrow \ell \bar{\nu}_\ell V_{DM}$]{\includegraphics[width=7.5cm]{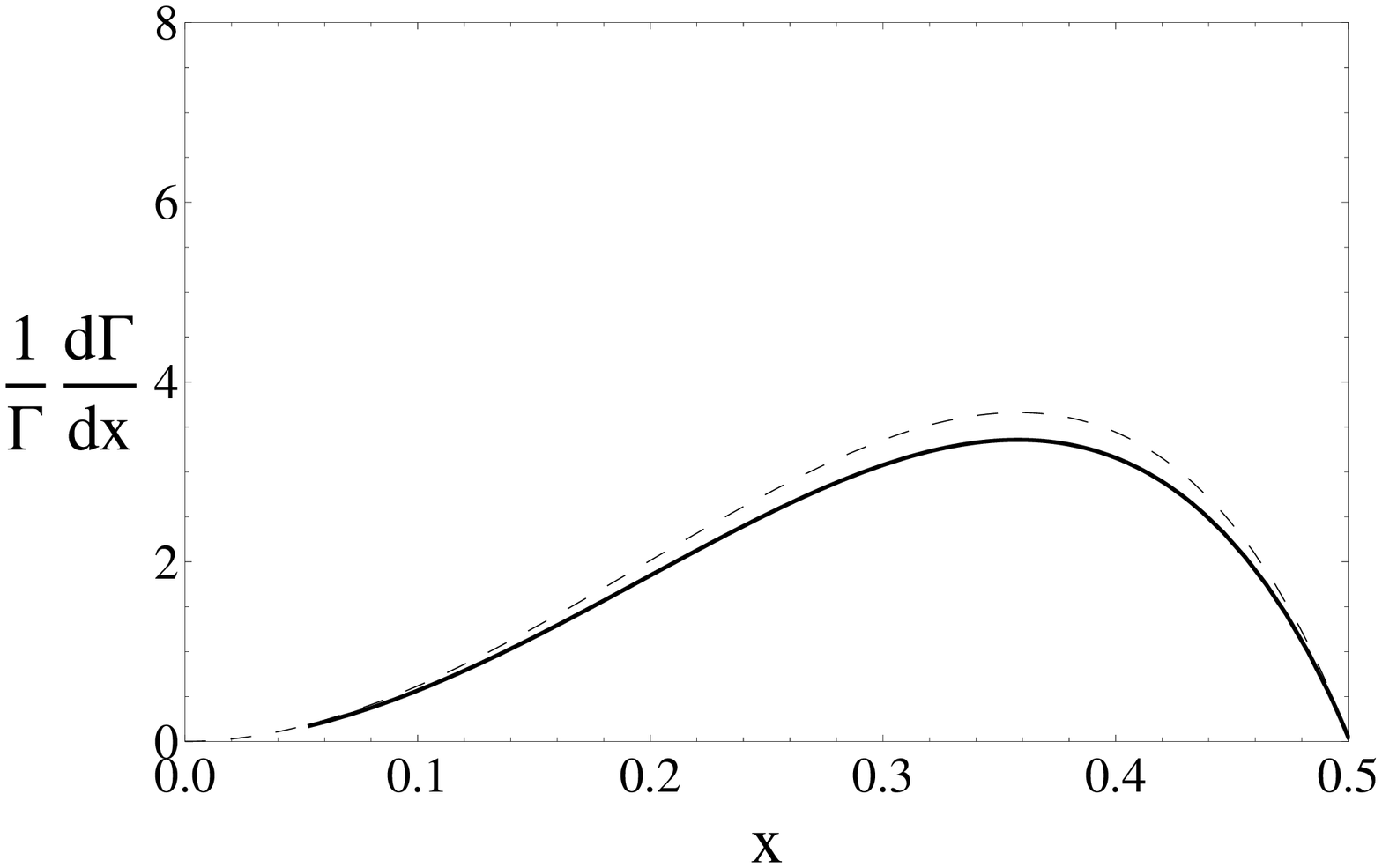}}
\end{center}
\caption{Normalized electron (dashed) and muon (solid) energy distributions for the  
heavy \{$(B^{\pm},D^{\pm},D_s^{\pm}$\} meson decay channels. Here $m_a=0$ and $x=E_\ell/m_B$.}
\label{FIG:LVDM_Lept_Dist} 
\end{figure}
Again, we fit the parameter $\kappa$ using the same data as in the axion-like DM case. The results are shown in 
Figure~\ref{FIG:kappaLVDM} where the $D^{\pm}\rightarrow \mu^{\pm} \bar{\nu}_\mu V$ decay can yield the best 
bound.
\begin{table}
\begin{center}
\footnotesize
\begin{tabular}{|c|c|c|}
\hline\hline
Channel &  \begin{tabular}{c} $\kappa^{-2} R_V (E_0)$\\ $E_0 = 100$~MeV\end{tabular} &  $ \kappa $ \\
\hline\hline
$\mathcal{B}\left(B^{\pm}\rightarrow \tau^{\pm} \bar{\nu}_\tau \right)$ & $ 8.8 \times 10^{-3} $ & $ \leq 11.6$  \\
$\mathcal{B}\left(D^{\pm}\rightarrow \mu^{\pm} \bar{\nu}_\mu \right)$ & $5.7 \times 10^{-1}  $ & $  \leq 0.31$\\
$\mathcal{B}\left(D_s^{\pm}\rightarrow \mu^{\pm} \bar{\nu}_\mu \right)$  & $ 5.4 \times 10^{-2}   $ & $ \leq 1.49$ \\
$\mathcal{B}\left(D_s^{\pm}\rightarrow \tau^{\pm} \bar{\nu}_\tau \right)$  &$1.3 \times 10^{-4}   $ & $ \leq  20.8$\\
\hline\hline 
$\mathcal{B}\left(B^{\pm}\rightarrow e^{\pm} \bar{\nu}_e \right)$ & $ 1.8 \times 10^{3} $ & $ \leq 11.2$  \\
$\mathcal{B}\left(B^{\pm}\rightarrow \mu^{\pm} \bar{\nu}_\mu \right)$ & $ 1.0 \times 10^{-1} $ & $ \leq 4.17$  \\
$\mathcal{B}\left(D^{\pm}\rightarrow e^{\pm} \bar{\nu}_e \right)$ & $ 1.5 \times 10^{3}  $ & $  \leq 0.83$\\
$\mathcal{B}\left(D^{\pm}\rightarrow \tau^{\pm} \bar{\nu}_\tau \right)$ & $ 1.8 \times 10^{-4}  $ & $  \leq 36.4$\\
$\mathcal{B}\left(D_s^{\pm}\rightarrow e^{\pm} \bar{\nu}_e	 \right)$  & $ 5.2 \times 10^{2}  $ & $ \leq 1.37$ \\
\hline\hline
\end{tabular}

\normalsize
\end{center}
\caption{Constraints on $\kappa$ using various decay channels. All other values are the same as in Table~\ref{FIG:BRALDM}. }
\label{FIG:kappaLVDM} 
\end{table}
Using the best constraint on $\kappa$ from the $D^{\pm}\rightarrow \mu^{\pm} \bar{\nu}_\mu V$ decay we can limit the 
contribution to yet-to-be-seen decays in Table~\ref{FIG:kLVDM2}.
\begin{table}
\begin{center}
\footnotesize
\begin{tabular}{|c|c|}
\hline\hline
Channel & $\mathcal{B}(\kappa = 0.31$) \\
\hline
\hline
$\mathcal{B}\left(B^{\pm}\rightarrow 
e^{\pm} \bar{\nu}_e \right)$ &$ 1.4 \times 10^{-9}  $\\
$\mathcal{B}\left(B^{\pm}\rightarrow 
\mu^{\pm} \bar{\nu}_\mu \right)$ &$3.6 \times 10^{-9} $ \\
$\mathcal{B}\left(D^{\pm}\rightarrow 
e^{\pm} \bar{\nu}_e \right)$ &$ 1.2 \times 10^{-6} $  \\
$\mathcal{B}\left(D^{\pm}\rightarrow 
\tau^{\pm} \bar{\nu}_\tau \right)$ &$ 1.7 \times 10^{-8} $ \\
$\mathcal{B}\left(D_s^{\pm}\rightarrow 
e^{\pm} \bar{\nu}_e \right)$ &$ 6.2 \times 10^{-6}  $\\
\hline\hline
\end{tabular}

\normalsize
\end{center}

\caption{Contributions to various yet-to-be-seen channels using the the fit on $\kappa$ in Table~\ref{FIG:kappaLVDM}. }
\label{FIG:kLVDM2} 
\end{table} 

As we can see, the constraints on the kinetic mixing parameter $\kappa$ are not very strong, but could be improved in the
next round of experiments at super-flavor factories.

\section{Conclusions}\label{Conclusions}

We considered constraints on the parameters of different types of bosonic super-WIMP dark matter from leptonic
decays of heavy mesons. The main idea rests with the fact that in the standard model the two-body leptonic decay 
width of a heavy meson $M=\{B^\pm, D^\pm, D_{s}^\pm \}$, or $\Gamma(M\rightarrow \ell\bar{\nu})$, is helicity-suppressed by 
$(m_\ell/m_B)^2$ due to the left-handed nature of weak interactions~\cite{Rosner:2010ak}. A similar three-body decay 
$M \rightarrow \ell\bar{\nu_\ell}X$ decay, which has similar experimental signature, is not helicity suppressed. 
We put constraints on the couplings of such DM particles to quarks. We note that the models of new physics 
considered here are very different from the models that are usually constrained in the new physics 
searches with leptonic decays of heavy mesons~\cite{Dobrescu:2008er}.

We would like to thank Andrew Blechman, Gil Paz, and Rob Harr for useful discussions.  
This work was supported in part by the U.S.\ National Science Foundation
CAREER Award PHY--0547794, and by the U.S.\ Department of Energy under Contract
DE-FG02-96ER41005. 




\end{document}